\documentclass[aps,prl,preprint,showpacs,superscriptaddress]{revtex4-1}
\usepackage{yfonts}
\usepackage{graphicx}
\usepackage{amsmath}
\usepackage{color}
\usepackage{epstopdf}
\usepackage{ulem}

\begin{document}

\title{Currentless reversal of N\'{e}el vector in antiferromagnets}
\author{Yuriy G. Semenov}
\affiliation{Department of Electrical and Computer Engineering,
North Carolina State University, Raleigh, NC 27695, USA}

\author{Xi-Lai Li}
\affiliation{Department of Electrical and Computer Engineering,
North Carolina State University, Raleigh, NC 27695, USA}

\author{Ki Wook Kim}
\affiliation{Department of Electrical and Computer Engineering,
North Carolina State University, Raleigh, NC 27695, USA}
\affiliation{Department of Physics, North Carolina State University,
Raleigh, NC 27695, USA}

\begin{abstract}
The bias driven perpendicular magnetic anisotropy is a magneto-electric
effect that can realize 90$^\circ$ magnetization rotation and even 180$%
^\circ $ flip along the easy axis in the ferromagnets with a minimal energy
consumption. This study theoretically demonstrates a similar phenomenon of
the N\'{e}el vector reversal via a short electrical pulse that can mediate
perpendicular magnetic anisotropy in the antiferromagnets. The analysis
based on the dynamical equations as well as the micro-magnetic simulations
reveals the important role of the inertial behavior in the antiferromagnets
that facilitates the N\'{e}el vector to overcome the barrier between two
free-energy minima of the bistable states along the easy axis. In contrast
to the ferromagnets, this N\'{e}el vector reversal does not accompany
angular moment transfer to the environment, leading to acceleration in the
dynamical response by a few orders of magnitude. Further, a small switching
energy requirement of a few attojoules illustrates an added advantage of the
phenomenon in low-power spintronic applications.
\end{abstract}

\pacs{75.75.-c, 75.78.Jp, 75.85.+t, 85.70.Ay}
\email{kwk@ncsu.edu}
\maketitle


In the early stages of spintronics, the antiferromagnets (AFMs) were
exploited almost exclusively in combination with free ferromagnetic layers.
The primary driver is the large magnetoresistance at these interfaces (i.e.,
the so-called giant magnetoresistance) that has since played a significant
role in the development of numerous applications such as the magnetic random
access memory \cite{Fert1988,Parkin1990}. Only recently have they been
recognized as an active spintronic medium with excellent dynamical
properties that can in fact claim advantages over the conventional
ferromagnetic counterparts \cite{Loktev2014,Cheng2015,Jungwirth2016}.
Similarly to the ferromagnets (FMs), the AFMs possess two quasistable states
along the easy axis that provide a natural system to encode or store the
binary information$-$the logical bit. However, the absence (or near absence)
of net magnetization can make its manipulation nontrivial, particularly with
external magnetic fields. An alternative approach for control is to take
advantage of the "effective" field or torque induced via the magnetic
interactions with adjacent layers whose materials are not necessarily
magnets.

One solution proposed earlier is in the manner of spin transfer torque (STT)
in FMs exploiting the dynamical origin of AFM magnetization \cite%
{McDonald2006,Gomonay2010,Cheng2015}. However, the current density needed to
generate sufficient torque remains high even for AFMs {\cite{Cheng2015}}.
Further, the weak magnetization tends to extend the transverse spin
decoherence length, requiring a thicker layer for the AFMs to rely on the
Slonczewski's mechanism of STT \cite{Slonczewski1996}. A potentially more
efficient approach may be possible via the electrostatic control of
perpendicular magnetic anisotropy (PMA). This effect has been demonstrated
in the numerous realizations of magnetoelectric heterostructures based on
the FMs, providing a highly potent means to achieve magnetization rotation
without involving any electrical current (see, for instance, Refs.~%
\onlinecite{Vaz2012,Hibino2015} as well as the references therein).


In this work, we theoretically explore the feasibility of PMA-mediated
switching between the two quasistable states in the AFMs. The investigation
is based on a mono-domain model of two compensated magnetic sublattices in
the Lagrangian approach. The main focus is on elucidating the basic physical
principles of the currentless N\'{e}el vector rotation rather than the
analysis of a particular implementation as there can be a wide range of
possibilities in the actual realization of the electrically controlled PMA
\cite{Vaz2012,Hibino2015,Roy2011,Semenov2012}. For one, the strain may be
used to affect the AFM anisotropy in analogy to FMs, while the specific
reports are yet to be available in the literature. The calculation clearly
illustrates the desired AFM switching by the temporal modulation of the PMA.
Further, the corresponding dynamical response is expected to much faster and
energy efficient than those of the FM counterparts.

The envision process is akin to the dynamical magnetization reversal that is
a well established procedure in the spin echo experiment via a $\pi $ pulse
in the rotating frame of reference \cite{Abragam1961}. Interestingly, a
similar concept has been extended to switch the nano-magnets \cite%
{Li2015,Grezes2016}. In the case of a FM, applying the PMA along the $z$
axis [$K_{A}(t)$] in the form of a single pulse can induce the effective
field $\mathbf{H}=\hat{\mathbf{z}}2m_{z}K_{A}(t)/M$, exerting a torque to
rotate $\mathbf{M}$ ($\mathbf{m}=\mathbf{M}/M$, $M=\left\vert \mathbf{M}%
\right\vert $) on the $x$-$y$ plane normal to the PMA (the blue curve in
Fig.~1) provided that the strength can overcome the in-plane axial
anisotropy. Unlike the magnetic resonance, $\mathbf{H}$ depends on the
instant state of $\mathbf{M}$ as shown above. Accordingly, the magnetization
executes a flip under the condition $2m_{z}\gamma M^{-1}\int
K_{A}(t)dt\simeq \pi $, where approximate conservation of $m_{z}$ is assumed
for the pulse duration and $\gamma $ is the gyromagnetic ratio. The
magnitude of $m_{z}$ can be controled by a weak external magnetic field \cite%
{Grezes2016}. Even thermal broadening of $m_{z}$ around the equilibrium
state $m_{z}=0$ evidently facilitates the magnetization switching at
sufficiently high temperature (i.e., a sizable non-zero $m_{z}$ component)
\cite{Berkov2002}.

At the first glance, a corresponding effect of PMA-induced reversal seems
infeasible in AFMs since the effective field cannot drive the N\'{e}el
vector $\mathbf{L}$ to precess around it (no net magnetization). Instead, $%
\mathbf{L} $ takes a short track to the redefined magnetic energy minimum
(i.e., along the $z$ axis) in a damped oscillatory behavior.
More precisely, the trajectory of the N\'{e}el vector is determined not only
by its instantaneous position $\mathbf{L}(t)$ but also by the velocity $\dot{{%
\mathbf{L}}} (t)$ (${\equiv } \frac{d}{dt}\mathbf{L}$). This means
that the vector $\mathbf{L}(t)$ tends to continue its path even
after the external driving field (i.e., the bias controlling the
PMA) is turned off. The underlying implication is that a properly
tailored $K_A(t)$, with the aid of the inertial motion, may realize
deterministic 180$^\circ$ inversion between two magnetic energy
minima of an AFM (the red curve in Fig.~1). Taking into account that
the AFM dynamics are exchange enhanced and not limited by
conservation of the angular momentum, the $\mathbf{L}$-vector
switching is expected to be much faster and require a significantly
smaller amount of energy than the FM counterparts.


In the analysis of the PMA influence on the AFM dynamics, a mono-domain
model of two compensated magnetic sublattices is solved by following the
Lagrangian approach developed earlier \cite{Baryakhtar1979,Andreev1980}.
This treatment conveniently allows the Lagrangian $\textswab{L}$ to be
expressed solely in terms of the AFM N\'{e}el vector $\mathbf{L}$ ($=\mathbf{%
M}_{1}-\mathbf{M}_{2}$) so long as the AFM magnetization $\mathbf{M}$ (=$%
\mathbf{M}_{1}+\mathbf{M}_{2}$) mediated by the misalignment of sublattice
magnetizations $\mathbf{M}_{1}$, $\mathbf{M}_{2}$ is relatively small.
Consequently, the length of the N\'{e}el vector $\left\vert \mathbf{L}%
\right\vert $ ($={M_{L}}\simeq \left\vert \mathbf{M}_{1}\right\vert
+\left\vert \mathbf{M}_{2}\right\vert $) can be approximately expressed as
an integral of the motion and the AFM magnetization acquires a dynamical
origin $\mathbf{M}=\frac{H_{ex}}{\gamma M_{L}}\mathbf{n}\times {\dot{\mathbf{%
n}}}$ at zero magnetic field, where $\mathbf{n=L/}{M_{L}}$ and $H_{ex}$ is
the exchange field acting between the sublattices \cite%
{Baryakhtar1979,Baryakhtar1980}.

At zero magnetic field, the Lagrangian
\begin{equation}
\textswab{L}=\frac{{M_{L}}^{2}}{2\omega _{ex}^{2}}{\dot{\mathbf{n}}}^{2}-W(%
\mathbf{n}).  \label{1}
\end{equation}%
determines the evolution of the AFM vector. Here, $\omega _{ex}^{2}=\gamma
^{2}H_{ex}{M_{L}}$ and $W(\mathbf{n})$ is the density of the anisotropy
energy, the magnitude of which can be dependent on the shape of the nano-magnet
as well as its interface characteristics \cite{Gomonay2014}. Combining
this inherent contribution with the electrically induced PMA along the $z$
axis, the total anisotropy can be expressed as
\begin{equation}
W(\mathbf{n,}t)=\frac{1}{2}%
\{K_{x}n_{x}^{2}+K_{y}n_{y}^{2}+[K_{z}+K_{A}(t)]n_{z}^{2}\},  \label{2}
\end{equation}%
where $K_{x}$, $K_{y}$ and $K_{z}$ are the values attributed
to the structure without external perturbation and $K_{A}(t)$ is the
electrically mediated PMA as defined earlier. For simplicity, the cubic and
higher-order terms are neglected in Eq.~(\ref{2}). Moreover, $K_{y}$ can be
set to zero without loss of generality when $\mathbf{n}^{2}=1$; this merely
amounts to the renormalization $K_{x}-K_{y}\rightarrow K_{x}$ and $%
K_{z}-K_{y}\rightarrow K_{z}$. Then, the magnetic relaxation toward the
local minimum of $W(\mathbf{n},t)$ can be incorporated into the kinetic
equation by way of a dissipation function
\begin{equation}
\textfrak{R}=\frac{\delta _{r}{M_{L}}^{2}}{2\omega _{ex}^{2}}{\dot{\mathbf{n}%
}}^{2},  \label{3}
\end{equation}%
which can be given in terms of the homogeneous line width $\delta _{r}$ of
AFM resonance. The correspondent Lagrange equation augmented with the
dissipation [Eq.~(\ref{3})] describes the evolution of the AFM vector in the
form of a Langevin second-order differential equation
\begin{equation}
\mathbf{n}\times \left[ \ddot{\mathbf{n}}+\omega _{ex}^{2}\frac{\partial }{%
\partial \mathbf{n}}\frac{W(\mathbf{n,}t)}{M_{L}^{2}}+\delta _{r}\dot{%
\mathbf{n}}\right] =0.  \label{5}
\end{equation}%
Similar expressions have been obtained earlier except $W(\mathbf{n,}t)$,
which now explicitly represents the time-dependent PMA {\cite{Gomonay2010}}.

To proceed further, it is convenient to represent Eq.~(\ref{5}) via polar
and azimuthal angles of vector $\mathbf{n}(t)=(\sin \theta \cos \varphi
,\sin \theta \sin \varphi ,\cos \theta )$ and introduce dimensionless time $%
t\rightarrow \omega _{r}t$ in terms of the zero-field AFM resonance
frequency $\omega _{r}=\sqrt{2\gamma ^{2}H_{ex}H_{an}}$. Here, $H_{an}$
represents the effective anisotropy field. Then, the corresponding
expressions take the form
\begin{equation}
\ddot{\theta}=\sin 2\theta \left[ \frac{1}{2}{\dot{\varphi}}^{2}+\xi
_{z}+\xi _{A}(t)-\xi _{x}\cos \theta \right] -\lambda \dot{\theta};
\label{t2}
\end{equation}%
\begin{equation}
\ddot{\varphi}\sin ^{2}\theta =-\dot{\theta}\dot{\varphi}\sin 2\theta +\xi
_{x}\sin ^{2}\theta \cos 2\varphi -\lambda \dot{\varphi}\sin ^{2}\theta ;
\label{t3}
\end{equation}%
where $\xi _{x}=K_{x}/M_{L}H_{an}$, $\xi _{z}=K_{z}/M_{L}H_{an}$, $\xi
_{A}(t)=K_{A}(t)/M_{L}H_{an}$, and $\lambda =\delta _{r}/\omega _{r}$.

To solve these coupled equations, appropriate initial conditions (defined as
$\theta _{0}$, $\varphi _{0}$, $\dot{\theta}_{0}$ and $\dot{\varphi}_{0}$
for the respective parameters) need to be specified. Note that the minimum
of the AFM anisotropy energy at $t=0$ (i.e., $\theta _{0}=\pi /2$ and $%
\varphi _{0}=0$) is just one particular realization among the possible
configurations at a finite temperature $T$. Similarly, the initial
"velocities" $\dot{\theta}_{0}$ and $\dot{\varphi}_{0}$ are also distributed
according to the "kinetic energy" with a dispersion around the thermal
energy $k_{B}T$. To account for all of the physically possible $\mathbf{n}%
(t) $ and $\dot{\mathbf{n}}(t)$, a distribution function $P(\mathbf{q})$ in
the phase space $\mathbf{q}$ [$=(\theta ,\varphi,\dot{\theta},\dot{\varphi})$%
] may be introduced in terms of the total magnetic energy $E$ of the AFM
with volume $V_{0}$. This quantity $E$ can be found directly from the
explicit form of the Lagrangian [Eq.~(\ref{1})] as
\begin{equation}
\frac{E}{V_{0}}=\dot{\mathbf{n}}\frac{\partial \textswab{L}}{\partial \dot{%
\mathbf{n}}}-\textswab{L}.  \label{6}
\end{equation}%
Then one can arrive, after some algebra, at the expression
\begin{equation}
E(\mathbf{q})=E_{M}\left( 4\dot{\theta}^{2}+4\dot{\varphi}^{2}\sin
^{2}\theta + \xi _{x}\sin ^{2}\theta \cos ^{2}\varphi + \xi _{z}\cos
^{2}\theta \right) ,  \label{7}
\end{equation}%
where $E_{M}=V_{0}M_{L}H_{an}$. Equation~(\ref{7}) explicitly defines $P(%
\mathbf{q})=N\exp \left[ -\frac{E(\mathbf{q})}{k_{B}T}\right] $ with a
normalization factor $N$; i.e., $\int P(\mathbf{q})d\mathbf{q}=1$. Then, the
range of typical initial conditions can be obtained in terms of the
root-mean-square value $\left\langle \Delta q_{i}\right\rangle =\sqrt{%
\overline{q_{i}^{2}}}$, where $\overline{q_{i}^{2}}=\int q_{i}^{2}P(\mathbf{q%
})d\mathbf{q}$. The problem is simplified when the relatively small
dispersion $\Delta \mathbf{q}$ [$=(\frac{\pi }{2}-\Delta \theta ,\Delta
\varphi ,\Delta \dot{\theta},\Delta \dot{\varphi})$] around the energy
extremum $\mathbf{q}_{0}=(\frac{\pi }{2},0,0,0)$ is taken into
consideration. The estimates give $\left\langle \Delta \dot{\theta}%
\right\rangle =\left\langle \Delta \dot{\varphi}\right\rangle =\sqrt{%
k_{B}T/8E_{M}}$, $\left\langle \Delta \theta \right\rangle =\sqrt{%
k_{B}T/2E_{M}|\xi _{x}-\xi _{z}|}$, $\left\langle \Delta \varphi
\right\rangle =\sqrt{k_{B}T/2E_{M}\xi _{x}}$. The increase of dispersion $%
\left\langle \Delta \theta \right\rangle $ with a reduction in the
difference $\xi _{x}-\xi _{z}$ is not surprising when considering that the $%
x $ axis ceases to be the easy axis as $\xi _{x}-\xi _{z}\rightarrow 0$.
Then, the $x$-$z$ plane instead becomes the easy plane with a much broader
initial distribution.

Now the solutions of the Eqs.~(\ref{t2}) and (\ref{t3}) can be obtained
under electrically induced PMA [i.e., $\xi _{A}(t)$] and initial conditions $%
\mathbf{q}(t=0)$ selected according to the thermal distribution $P(\mathbf{q}%
)$. For the numerical results, we exploit the simplest case of easy axis AFM
assuming $K_{x}=-2.5\cdot 10^{5}$ erg/cm$^{3}$, $K_{z}=0$ and adopt the
typical AFM zero-field resonance frequency $f_{r}(=\omega _{r}/2\pi )$ of $%
180$ GHz at the sublattice magnetization $M_{L}/2$ of 200 Oe. These
parameters correspond to the effective fields $H_{ex}=270$ T and $H_{an}=800$
Oe. The quantity of the magnetic energy $E_{M}$ is linearly proportional to
the volume $V_{0}$ (assumed to be 60$\times $60$\times $2 nm$^{3}$) whose
magnitude would provide nonvolatility at room temperature ($\approx $ 40$%
k_{B}T$). The PMA in the form of a rectangular pulse with amplitude $K_{A}=${%
$-4\times 10^{5}$ erg/cm$^{3}$} and duration $\Delta t$ is assumed at $t=0$
that alters the easy axis to be essentially along the $z$ direction. Thus
the full set of the parameters $\xi _{x}=-0.5$, $\xi _{A}(t)=-0.8$ [$t\in
(0,\Delta t)$], $\xi _{A}(t)=0$ [$t\notin (0,\Delta t)$], and damping factor
$\lambda =0.4$ determines the N\'{e}el vector evolution in terms of the Eqs (%
\ref{t2},\ref{t3}) and dimensionless time $t\omega _{L}$. The
corresponding
thermal broadening of the initial states around the energy minimum $E(%
\mathbf{q}_{0})$ is estimated to be $\left\langle \Delta \varphi
\right\rangle \simeq \left\langle \Delta \theta \right\rangle \simeq
5^{\circ }$, $\left\langle \Delta \dot{\theta}\right\rangle \simeq
\left\langle \Delta \dot{\varphi}\right\rangle \simeq 0.06$ that will be
used in following calculations.

Figure~2(a) clearly illustrates the pendulum-like dynamics of AFM vector.
Shifting of potential minimum along with the PMA exerts the N\'{e}el vector
moving to a new equilibrium state. A very short perturbation in form of PMA
pulse may only lead to a minor deviation from starting point and following $%
\mathbf{L}$ relaxation to initial state (curves 1). Relatively long pulse or
stationary PMA leads to the common PMA effect of $90^{\circ }$ turn after
several $\mathbf{L}$ oscillations around a new minimum (curves 2). These
vibrations around neutral position suggest to explore the effect of
intermediate pulse durations $\Delta t$. If the PMA is interrupted when the N%
\'{e}el vector reaches the vicinity of $z-$direction and continua moving in
the line of reversal state $n_{x}\equiv L_{x}/\left\vert \mathbf{L}%
\right\vert \simeq -1$ away from initial state $n_{x}=+1$, it will appear in
the attractive zone of the equilibrium state with $n_{x}=-1$. Then its
following relaxation results in deterministic $\mathbf{L}$-vector switch
[Fig. 2(b), curves 3]. Such behavior can be observed for $\Delta t=4-7$ ps.
However prolonging of the PMA pulse reverses the direction of moving so that
backward pass of the $90^{\circ }$ extremum will return $\mathbf{L}$ to the
initial attractive zone (curves 4). Apparently, extension of pulse duration
leads to the "kinetic energy" dumping that increases the role of thermal
fluctuations with a random selection between $+x$ and $-x$ directions.
Therefore the proper $\Delta t$ selection offers the means of deterministic
180$^{\circ }$ reversal.

It would be instructive to compare the AFM dynamics based on the starting
monodomain approach [Eqs.~(\ref{t2}) and (\ref{t3})] with simulation of the
AFM switch in terms of micro-magnetic approach. The latter represents an AFM
as the foliated FM layers with AFM interaction between them. In turn, each
FM layer consists of small FM cells driven by local exchange fields
according to Landau-Lifshitz-Gilbert equation \cite{MicroMagnet}. For
particular simulation we choose the FM cell sizes 0.5$\times $0.5$\times $%
0.5 nm$^{3}$ and assign the previously used magnetization and anisotropy.
The inter (intra) layer exchange constants $J=-(+)5\times 10^{-7}$ erg/cm
suppose to determine the AFM resonance frequency $180$ GHz applied in
monodomain approximation. The Gilbert damping parameter $\alpha =\lambda
\sqrt{2H_{an}/H_{ex}}=0.01$ evokes attenuation of $\mathbf{L}$ vector
oscillations. Besides the initial states of magnetizations assume to be same
for each FM layer (i.e. $\pm $5$^{\circ }$ away from extremum points) so
that the total magnetization $M=0$ at $t=0$ (Fig. 3a). As it was mentioned,
such a state corresponds to zero "velocity" $\dot{\mathbf{L}}=0$ (Fig.3).
This starting point results in some difference in N\'{e}el vector dynamics
compared with calculations depicted at Fig. 2. It is remarkable that despite
of micro-magnetic simulation allows non-coherent behavior of FM cells, both
approaches demonstrate similar behaviors. They interactively demonstrate the
quick N\'{e}el vector switch along the trace escaping a pass through the $y$%
-axis that is unavoidable for magnetization switch in the FM (Fig.~1).

As soon as PMA pulse tailoring is a crucial circumstance to reach the
desirable effect of AFM vector switch (Fig. 2), we estimate the conditions
of successful device performance in terms of the strength and duration of
PMA pulses.   Fig. 4 shows the correspondent phase diagram for AFM
parameters used for Fig.2 at various pulse durations and amplitudes. The
darker (blue) and lighter (green) region represent the final L-vector
equilibrium states with $n_{x}=1$ (returning back to initial state) and $%
n_{x}=-1$ (success reversal) respectively. The alternative property
of the phase diagram stems from the oscillatory behavior of the
pendulum-like AFM dynamics. Note that the thermal fluctuations may
become a source of uncertainty at the long pulse duration. In such a
case the velocity damping diminishes the "kinetic energy" to the
thermal limit or even below so that thermal fluctuations randomize
the final states. Thus in our case of damping parameter $\lambda
=0.4$, only the first or second reversal region could ensure the
required reversal probability. Apparently the practical
implementation of the particular AFM among the large number of the
available magnetic materials would rescale the graphs depicted at
Figures 2, 3 and 4. However the qualitative properties of
pendulum-like dynamics makes the robust switching effect under the
properly tailoring of the PMA pulse.

In summary, an effective mechanism of AFM vector switching is proposed. In
contrast to STT in AFM \cite{McDonald2006,Gomonay2010,Cheng2015} the
PMA-mediated $\mathbf{L}$ reversal occurs in an electric field without
high-density electric current. As such, the energy consumption for actual
device operation can be expected in the range of few aJ \cite{Xiaopeng2014}.
As to practical implementation, one should provide an infallible method of
discriminating of the two metastable $\mathbf{L}$ directions. The GMR in the
structure that consists of FM with fixed magnetization direction and
adjacent free AFM may resolve this problem \cite{Fert1988,Parkin1990}. A
different approach can be rely on surface conductance of topological
insulator, which is sensitive to magnetization direction of proximate layer $%
\mathbf{M}_{1}$ or $\mathbf{M}_{2}$ of AFM. Apart from the evident
application in energy saving fast memory cells \cite{Kos2006} the extremely
strong non-linearity of the response on input signal would offer the
applications in logic devices. Indeed, the relatively weak input signals (as
a logic "1") may not solely switch $\mathbf{L}$ direction, but combine both
inputs would successfully switch $\mathbf{L}$ vector realizing logic routine
"AND". Similarly, stronger input signals and proper their combination could
realize operation "OR". Thus the magneto-electric structures with electrical
control of PMA in AFM offer the new capabilities of spintronic devices that
would excel the CMOS counterparts in speed and efficiency.

This work was supported, in part, by the US Army Research Office and FAME
(one of six centers of STARnet, a SRC program sponsored by MARCO and DARPA).

\clearpage

\begin{center}
\begin{figure}[tbp]
\includegraphics[width=5cm,angle=0]{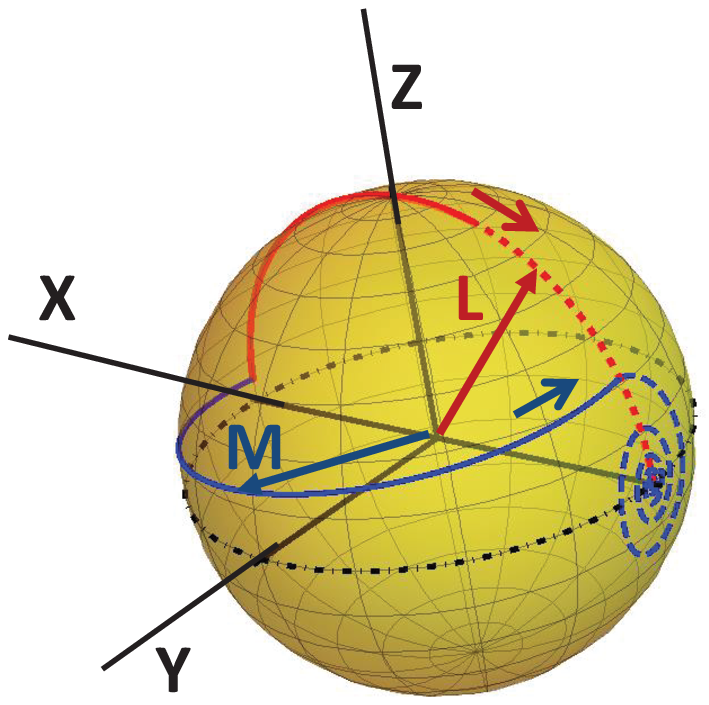}
\caption{(Color online) Schematic illustration of the traces of FM
magnetization vector $\mathbf{m}=\mathbf{M}/M$ and AFM vector $\mathbf{n}=%
\mathbf{L}/L$ reversal induced by the exposure of PMA-pulses. Small
deviation from extremal point $\mathbf{m}=\mathbf{n=}(1,0,0)$ exerts the $%
\mathbf{m}$ rotation along the track (solid blue curve) close to equatorial $%
x-y$ plane delineated by dot-dashed circle on Bloch sphere. After PMA $%
\protect\pi -$pulse terminating the $\mathbf{m}$ relaxes in the direction of
reversal state $(-1,0,0)$ (dashed blue line). Much shorter PMA pulse exerts
the $\mathbf{n}-$ relaxation into zenith direction (solid red line), which
is passed along-track direction due to accumulated velocity. After PMA
attenuates the $\mathbf{n}$ continues to relax into reversal state along
easy axis (dashed red line).}
\end{figure}
\end{center}

\clearpage

\begin{center}
\begin{figure}[tbp]
\includegraphics[width=7.5cm,angle=0]{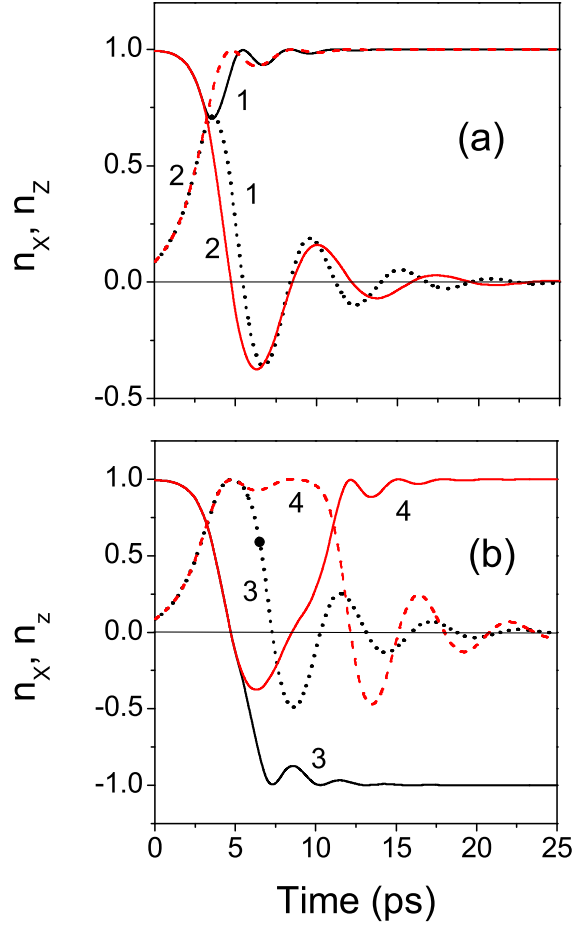}
\caption{(Color online) The evolution of $x-$ and $z-$ components of AFM
vector $\mathbf{n=L/}L $ under PMA pulses of (a) short $\Delta t=3$ ps
(curves 1) and long $\Delta t=25$ ps (curves 2) durations. (b) The $\mathbf{n%
}$ response on intermediate PMA pulse durations $\Delta t=6$ ps (curves 3)
and $\Delta t=9$ ps (curves 4). Solid (dashed) lines represent $n_{x}(t)$ [$%
n_{z}(t)$]. The $n_{y}(t)$ changes insufficiently because of hard $y-$ axis
and is not shown. Calculations were carried out in terms of Eqs.~(  \protect
\ref{t2}) and (\protect\ref{t3}) for parameters listed in the main text.}
\end{figure}
\end{center}

\clearpage

\begin{center}
\begin{figure}[tbp]
\includegraphics[width=7.5cm,angle=0]{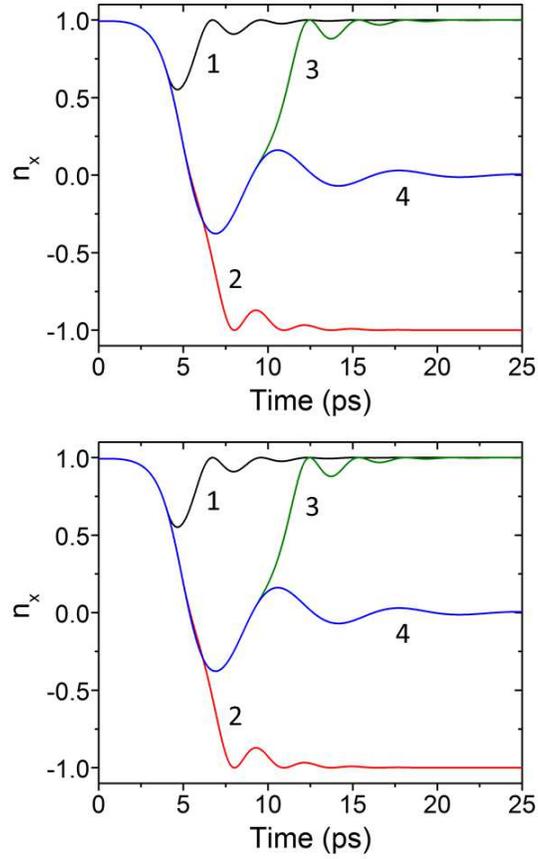}
\caption{(Color online) Comparison of the N\'{e}el vector dynamics obtained
with (a) the micro-magnetic simulations and (b) the monodomain approach. The
durations 3 ps (curve 1), 5 ps (curve 2), 9 ps (curve 3) and 25 ps (curve 4)
of PMA-pulses represent the qualitatively different responses of AFM vector
in both approaches. The material parameters are discussed in the main text.}
\end{figure}

\begin{figure}[tbp]
\includegraphics[width=7.5cm,angle=0]{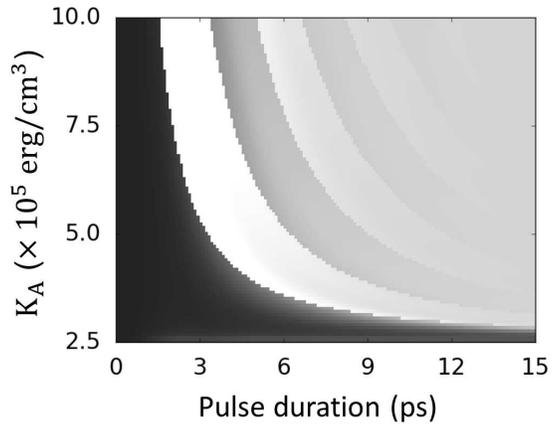}
\caption{The target states of N\'{e}el vector of easy-axis AFM at
different PMA pulse duration and its strength. The darker and
lighter region represent the starting (initial) state with $n_x=1$
and reversal state with $n_x=-1$ respectively. The darkness is
weighted by the total magnetic energy damping with damping
coefficient $\alpha=0.01$. The higher contrast region would have
higher probability of deterministic switching, while the lower
contrast region (top right) would be vulnerable to thermal noise.}
\end{figure}
\end{center}

\end{document}